\documentclass{article}
\usepackage{amssymb}
\usepackage{graphicx}
 \newcommand{\singlefig}{.75\textwidth}
 \newcommand{\doublefig}{\textwidth}
\date{June 12, 2003}
\title
{\bf Multi-site H-bridge breathers in a DNA--shaped double strand}
\author{{\bf D. Hennig}\\
Freie Universit\"{a}t Berlin, Fachbereich Physik\\
 Institut f\"{u}r Theoretische Physik\\
Arnimallee 14, 14195 Berlin, Germany \and {\bf J.F.R. Archilla}
\thanks{Corresponding author, Email: archilla@us.es}
\\Group of Nonlinear Physics of the University of Sevilla.\\ Departamento de F\'{\i}sica
Aplicada I, E.T.S.I. Inform\'atica. \\Avda Reina Mercedes s/n,
41012--Sevilla, Spain }
\begin{document}

\maketitle


PACS numbers: 87.-15.v, 63.20.Kr, 63.20.Ry \\

\begin{abstract}
\noindent We investigate the formation process of nonlinear
vibrational modes representing broad H-bridge multi--site
breathers in a DNA--shaped double strand.
 Within
a network model of the double helix we take individual motions of
the bases within the base pair plane into account. The resulting
H-bridge deformations may be asymmetric with respect to the helix
axis. Furthermore  the covalent bonds
 may be deformed distinctly in the two backbone strands.
 Unlike other authors
that add different extra terms we limit the interaction to the
hydrogen bonds within each base pair and the covalent bonds along
each strand. In this way we intend to make apparent the effect of
the characteristic helicoidal structure of DNA. We study the
energy exchange processes related with the relaxation dynamics
from a non-equilibrium conformation. It is demonstrated that the
twist-opening relaxation dynamics of a radially distorted double
helix attains  an equilibrium regime characterized by a multi-site
H-bridge breather.

\end{abstract}


\section{Introduction}

Studies of chemical and physical properties of DNA have attracted
considerable interest among physicists and biologists because of
their relevance for a variety of biological processes, such as DNA
transcription, gene expression and regulation and DNA replication
\cite{Stryer}. In the context of the transcription
process, the coding sequence on a DNA strand has to be made
accessible to the RNA polymerase which necessitates that hydrogen
bonds connecting the two strands have to be (temporarily) broken
so that the DNA strands separate. This melted segment of DNA
encloses $15-20$ opened base pairs and is called the transcription
bubble.

The opening of the DNA molecule is a very complex process and a
comprehensive explanation of the actual mechanisms underlying the
observed dynamical processes is still to come. To tackle the
problem of DNA dynamics nonlinear models of the DNA double helix
have been proposed during the past two decades
\cite{PB}-\cite{BCP}. Different from molecular dynamics simulation
methods, for which a great deal on the detailed reproduction of
the molecular DNA structure is spent, the nonlinear models focus
on the most relevant structural features only.  The strength of
the rather abstract nonlinear models of DNA lies then in the
straightforward manner in which their soliton and breather
solutions reproduce the prevalent dynamical behavior related with
the strong energy localization and stable energy transport
observed experimentally during the transcription process.

To model the base pair opening in DNA,  Peyrard and Bishop (PB)
proposed a planar ladder-like model of DNA assigning each base
pair a vertical inter-strand vibrational degree of freedom
simulating the stretchings and compressions of the corresponding
H-bridges \cite{PB}. The binding forces of the hydrogen bridges
are described by a Morse potential. The bases itself are treated
as point masses. Horizontally, the bases on the same strand are
coupled via the
 covalent
 interaction described by harmonic potentials. The PB has
been extensively studied and localized oscillating solutions
(breathers) and moving localized excitations \cite{Lipniacki} have
been found reflecting successfully some typical features of the
DNA opening dynamics such as the magnitude of the amplitudes and
the time scale of the breathing of the 'bubble' occurring prior to
thermal denaturation \cite{PB}. In addition, studies in the
context of thermal denaturation have been based on the PB model
incorporating a heat bath \cite{Dauxois}-\cite{Theodorakopoulos}.

Typical for many biomolecules is their interplay between
functional processes and structural transitions \cite{Stryer}. In
this respect the transcription process relying on the opening of
DNA represents a prominent example. Indeed, it has been found that
the bubble formation  is strongly correlated with twist
deformations and local openings are always connected with a local
untwist of the double helix \cite{BCP},\cite{Barbithesis},\cite{biophys}. In
order to account for the helicoidal structure of DNA the PB model
has been significantly extended by Barbi, Cocco and Peyrard (BCP)
\cite{BCP}. In their model of a more realistic description of DNA,
two degrees of freedom per base pair are introduced. There is  a
radial variable measuring the distance between  two H-bridged
bases along a line that connects them in the base plane being
perpendicular to the helix axis. Further, the twist angle between
this connecting line and a reference direction determines the
orientation of the H-bridge. The twist-opening dynamics is
described by  radial breather solutions combined with kink-like
solutions in the angular variable which have been constructed with
the help of a multiple scale expansion technique \cite{BCP}.

In the current study of the nonlinear dynamics related with the
opening process in DNA our aim is twofold. First of all, we study
the energy exchange processes and the relaxation dynamics in DNA
molecules after their excitation into a non-equilibrium
conformation. Such an investigation is associated with recent
mechanical experiments performed with single  DNA molecules
\cite{Marko}-\cite{Clausen} forced away from their equilibrium
conformations. After force applications energy redistribution
within the DNA molecule takes place such that a new equilibrium
conformation is attained \cite{Crisona}.
Furthermore, for strong enough radial forces the mechanical
unzipping of single DNA molecules is achievable \cite{Bockelmann}.
We remark that for the current study our assumption is that the
DNA molecules are imposed to not too strong external radial forces
excluding the (immediate) separation of the strands. Secondly,
besides the relaxation dynamics within B-DNA whose two strands are
supposed to have been pulled apart in a certain region, we direct
our attention on the formation of oscillating bubbles occurring
previous to denaturation. In particular, within a nonlinear model
approach we aim for the creation of multi-site radial breathers,
which, compared with the one-site breather solutions
obtained in \cite{BCP},\cite{biophys},
 provide stronger opening of an fairly extended segment of the DNA
reproducing the experimentally observed
oscillating bubbles involving $15-20$ base pairs.

The structure of the double helix of B-DNA is modeled by a steric
network of oscillators in the frame of the base pair picture
\cite{PB},\cite{BCP} taking into account deformations of the
hydrogen bond within a base pair and twist motions between
adjacent base pairs. In augmentation of the oscillator model for
the helicoidal DNA structure introduced in \cite{BCP} we allow for
individual motions of the bases within the base pair plane such
that the vibrations of the H-bridges are no longer exclusively
symmetric with respect to the helix axis unlike the symmetric
radial motions considered in \cite{BCP}.

It is worth mentioning that when constructing helical variants of
the Peyrard--Bishop model, the description of the hydrogen-bridge
and covalent bonds is fairly clear: an asymmetric soft potential
for the hydrogen bonds and harmonic terms for the covalent bond,
in the hypothesis that the latter are more rigid and therefore the
oscillations are small. It is not so evident how to model the
stacking interaction in a straightforward manner. The physical
origin of it lies in the bonds between overlapping bases, but
 these do
 not appear explicitly in the model. Different heuristic
solutions have been tried as it will be commented in more detail
later with interesting results. Here we focus on other aspects of
the problem and take the helical shape of the DNA as given and add
no extra term to maintain it. In this way we pretend to explore
the effect of the helical shape in itself. The results are
remarkable, the helical shape provides a change of the type of the
effective coupling  bringing about the possibility of stable
multi--site H-bridge breathers, which have to be differentiated
from broad one-site breathers. In this way much larger openings of
the double strand are possible.

The paper is organized as follows: In the second section we
describe our extended steric network model for the
 helical
 structure
of the double helix of B-DNA. The third section deals with the
relaxation dynamics within B-DNA molecules
forced into locally distorted configurations  and the fourth section is devoted to the
construction of multi-site radial breathers.  In the fifth section we discuss
the linear modes of the breathers and their relation to the spectral features of the localized
solutions.  Finally, we present our conclusions.

\section{Oscillator model for the helical structure of DNA}

In  our DNA model we incorporate the basic geometrical features of
the DNA double helix structure which are essential to model the
nonlinear dynamics of the twist-opening process.
Similar to the approach in \cite{BCP}, we treat the right-handed
helical form of B-DNA, which is a polymeric molecule composed of
two coiled strands of nucleotides forming a double helix, as a
 helical
 double-stranded oscillator system. The constituents of the
latter represent the nucleotides which are regarded as single
nondeformable entities. Thus, no inner dynamical degrees of
freedom of the nucleotides are taken into account which is
justified by the time scale separation between the small-amplitude
and fast vibrational motions of the individual atoms and the
slower and relatively large-amplitude motions of the atom groups
constituting the nucleotides \cite{Stryer}. Concerning the
structural components, each nucleotide is composed of a sugar, a
phosphate and a base. The sugar-phosphate groups of neighboring
nucleotides on the same strand are linked via covalent bonds
establishing the interaction related with the rigid backbone to
the strand. There is a base attached to every sugar. Since, for
simplicity, we do not distinguish between the four different types
of bases, the nucleotides are considered as identical objects of
fixed mass. Two bases on opposite strands are linked via hydrogen
bonds holding the two strands of DNA together.

In the BCP twist-opening model, based on the base pair picture of
DNA, the helicoidal structure of DNA has been conveniently
described in a cylindrical reference system where each base pair
possesses two degrees of freedom, namely a radial variable
measuring the distance between a base and the central helix axis
(viz., deformations of the corresponding H-bridge) and the angle
with a reference axis in a plane perpendicular to the central axis
which defines the twist of the helix \cite{BCP}.

Our model approach of the structural dynamics of the helicoidal
DNA models is inspired by the ones used in
Ref.~\cite{Barbithesis}--\cite{Agarwal}, but differs from them in
two important points: a)~We release the constraint that the two
bases in a base pair perform solely  vibrational variations of the
hydrogen bond length symmetric to the central axis along a line
which connects the two bases crossing the central helix axis;
b)~Those models introduce for convenience extra potential terms
which  we discard because of the following reasons:
 On the one
hand, not only they are sufficiently justified on physical
grounds. On the other hand, we are interested in the effect of the
helical structure on breather formation and try to keep the  model
as simple as possible to avoid masking the dynamical behavior.
These differences will bring about remarkable consequences on
multibreather stability and on the phonon spectrum as will be
shown in the next sections. We also neglect vertical movements as
Ref.~\cite{Agarwal} shows that they are not significant.
 Therefore, we extend
 the BCP model in the sense that we
treat  the two H-bridged strands individually and take
individual two-dimensional equilibrium positions as well as
displacement coordinates of a base within its base pair plane into
account. As a consequence, the line connecting the two bases of a base
pair, which is supposed to coincide with the orientation of the
hydrogen bridge, does not necessarily intersects the central axis
as distinct from the BCP model. In other words, for the
equilibrium configuration of the double helix  the two bases of a
base pair may  be positioned asymmetrically in radial
direction and may be rotated also by different angles around the helix axis
  allowing for
the simulation of polymorphic helical DNA matrices deviating from
the perfectly regular equilibrium helix structure. In fact, real
DNA molecules exhibit random structural imperfections of their
 equilibrium double helix caused by external and internal influencing factors
 such as the distorting impact of the solvent environs. Moreover,
 with our approach the DNA lattice dynamics can be initialized with
 arbitrary  excitation patterns connected with
 individual displacements of each base
different from the symmetric H-bridge deformations discussed
previously in \cite{BCP}.

\begin{figure}
  \begin{center}
    \includegraphics[width=\singlefig]{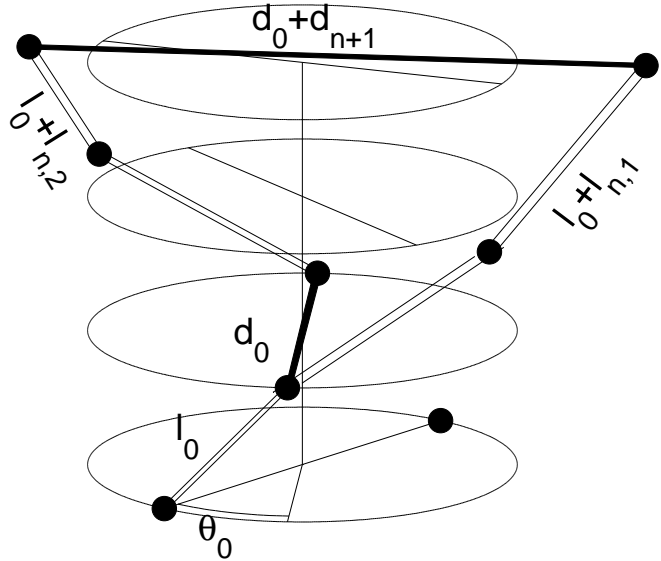}
  \caption{Schematic representation of the
helicoidal structure of the DNA model.} \label{fig:fig1}
\end{center}
\end{figure}

 We describe the double helix structure in a Cartesian coordinate
system whose $z-$axis corresponds to the central helix axis as
sketched in Fig.~\ref{fig:fig1}. The base pairs are situated in
planes perpendicular to the central helix axis and the vertical
distance between two consecutive planes is given by $h$. For the
equilibrium configuration of B-DNA each base, with equilibrium
coordinates $x_{n,i}^{(0)}$ and $y_{n,i}^{(0)}$, is rotated around
the central axis by an angle $\theta_{n,i}^{(0)}$. The index pair
$(n,i)$ labels the $n$-th base on the $i-$th strand with $i=1,2$
and  $1\leqq n\leqq N$\,, where $N$ is the number of base pairs
considered.
 The equilibrium distance between two
bases within a base pair, $d_0$, is determined by
\begin{equation}
d_0=\sqrt{\left(\,d_n^x\,\right)^2\,+\,
\left(\,d_n^y\,\right)^2}\,,
\end{equation}
where  $d_n^x=\,x_{n,1}^{(0)}-x_{n,2}^{(0)}$ and
$d_n^y=\,y_{n,1}^{(0)}-y_{n,2}^{(0)}$ are the projections of the
line connecting the two bases on the $x,y-$axes of the coordinate
system. The deviations $d_n$ from $d_0$ through displacements, $x_{n,i}$
and $y_{n,i}$, of the bases from their equilibrium positions,
$x_{n,i}^{(0)}$ and $y_{n,i}^{(0)}$, are expressed as
\begin{equation}
d_n=\sqrt{\left(\,d_n^x+x_{n,1}-x_{n,2}\right)^2\,+\,
\left(\,d_n^y+y_{n,1}-y_{n,2}\right)^2}-d_0\,.
\label{eq:dn}
\end{equation}
For later use we introduce the quantity
\begin{equation}
\theta_n=\arctan\frac{d_n^y+y_{n,1}-y_{n,2}}{d_n^y+x_{n,1}-x_{n,2}}+2\,m\,\pi\,,
\label{eq:thetan}
\end{equation}
 as the angle
between the $x-$axis (as the reference direction) and the line
connecting two (displaced) bases of a base pair measuring the
alignment of the related H-bridge, $m$ being a integer to assure
the monotonicity of $\theta_n$ with respect to $n$.

The three-dimensional equilibrium distance between two adjacent
bases on the same strand is given by
\begin{equation}
l_0=\sqrt{h^2+\left(\,x_{n,i}^{(0)}-x_{n-1,i}^{(0)}\right)^2\,+\,
\left(\,y_{n,i}^{(0)}-y_{n-1,i}^{(0)}\right)^2}\,,
\end{equation}
and deviations from $l_0$ are determined by
\begin{equation}
l_{n,i}=\sqrt{h^2+\left(\,L_{n,i}^x+x_{n,i}-x_{n-1,i}\right)^2\,+\,
\left(\,L_{n,i}^y+y_{n,i}-y_{n-1,i}\right)^2}-l_0\,,
\end{equation}
with $L_{n,i}^x=x_{n,i}^{(0)}-x_{n-1,i}^{(0)}$ and  $L_{n,i}^y=y_{n,i}^{(0)}-y_{n-1,i}^{(0)}$.
The Hamiltonian of our model is then of the form
\begin{equation}
H=E_{kin}+U_{h}+U_{c}\,,\label{eq:Hdna}
\end{equation}
where $U_{h}$ and $U_{c}$ represent the potential energy part for
the H-bond and the covalent bonds, respectively. The kinetic
energy is determined by
\begin{equation}
E_{kin}=\frac{1}{2m}\,\sum_{i=1,2}\,
\sum_{n=1}^{N}\,\left[\,\left(p_{n,i}^{(x)}\right)^2
+\left(p_{n,i}^{(y)}\right)^2\right]\,,\label{eq:Ekin}
\end{equation}
where $m$ is the mass of a base and $p_{n,i}^{(x,y)}$ denote the
$(x,y)-$component of the momentum of a base.

The vibronic potential $U_{h}$ represents the deformation energy
of the hydrogen bonds linking the two bases of the pair. The
dynamical deviations from the equilibrium bond length $d_n(t)$ are
supposed to evolve in a Morse potential and the potential energy
related to the displaced hydrogen bonds is given by
\begin{equation}
U_{h}= D\,\sum_{n=1}^{N}\,
\left[\,\exp\left(-\alpha\, d_n\right)\,-1\,\right]^2\,,
\label{eq:Uhyd}
\end{equation}
$\alpha^{-1}$ and $D$ are respectively the width and the depth of  the Morse potential well, corresponding the latter to the dissociation energy of the H--bond. In our simplified model of the DNA double helix we do
not distinguish between the two different pairings in DNA, namely
the G-C and the A-T pairs. The former pair involves three hydrogen
bonds while the latter involves only two.

Compared to the weak and flexible hydrogen bonds (with bond
energies of the order of $0.04-0.3\,eV$) the covalent bonds
between the sugar-phosphate groups of neighboring nucleotides on a
strand are rather strong and rigid (with bond energies of the
order of $2-10\,eV$). Therefore it is appropriate to treat the
potential of the covalent bonds, simulated by elastic rods, in the
harmonic approximation given by
\begin{equation}
U_{c}=K\,\sum_{i=1,2}\,\sum_{n=1}^{N}\,l_{n,i}^2\,,\label{eq:Uback}
\end{equation}
where $K$ is the elasticity coefficient.

With regard to other dynamical degrees of freedom we remark that
longitudinal acoustic motions along the strands are significantly
restrained by the rigidity of the phosphate backbone \cite{Ye}.
Therefore we discard displacements of the bases in $z-$direction.
Hence, the structural dynamics of base motions is restricted to
the base planes.

We will use the values of the parameters for B-DNA molecules given
by \cite{Stryer},\cite{BCP},\cite{Barbithesis}:
$\alpha=4.45$\,\AA$^{-1}$, $r_0 \thickapprox 10$\,\AA,
$h=3.4$\,\AA\,, $\theta_0=36^\circ$, $D=0.04\,\mathrm{eV}$,
$K=1.0\,\mathrm{eV}$\,\AA$^{-2}$ and
$M=300\,\mathrm{amu}\,=\,4.982\times10^{-25}\mathrm{kg}$.

With a suitable time scaling  $t\rightarrow \sqrt{D
\alpha^2/m}\,t$ one passes to a dimensionless formulation with
quantities:
\begin{eqnarray}
 \tilde{x}_{n,i}&=&\alpha
x_{n,i}\,, \,\,\,\, \tilde{y}_{n,i}=\alpha y_{n,i}\,,\,\,\,\,
\tilde{p}_{n,i}^{(x)}=\frac{p_{n,i}^{(x)}}{\sqrt{mD}}\,,\,\,\,\,
\tilde{p}_{n,i}^{(y)}=\frac{p_{n,i}^{(y)}}{\sqrt{mD}}\,,\\
 \tilde{K}&=&\frac{K}{\alpha^2 D}\,,\,\,\,\,
\tilde{d}_{n}=\alpha\, d_{n}\,,\,\,\,\,
\tilde{r}_{0}=\alpha\,r_{0}\,,\,\,\,\,\tilde{h}=\alpha\,h \,.
\end{eqnarray}
In the following, we omit the tildes.

The equations of motion are derived from the Hamiltonian
(\ref{eq:Hdna}) and read as
\begin{eqnarray}
\dot{x}_{n,i}&=&p_{n,i}^{(x)}\,,\label{eq:xdot}\\
\dot{p}_{n,i}^{(x)}&=&2\left[\,\exp(-d_n)-1\,\right]\,\exp(-d_n)\,
\frac{\partial d_n}{\partial x_{n,i}}\nonumber\\
&-&2K\left[\,l_{n,i}\frac{\partial l_{n,i}}{\partial x_{n,i}}+
l_{n+1,i}\frac{\partial l_{n+1,i}}{\partial x_{n,i}}\,\right]\,,\\
\dot{y}_{n,i}&=&p_{n,i}^{(y)}\,,\label{eq:ydot}\\
\dot{p}_{n,i}^{(y)}&=&2\left[\,\exp(-d_n)-1\,\right]\,\exp(-d_n)\,
\frac{\partial d_n}{\partial y_{n,i}}\nonumber\\
&-&2K\left[\,l_{n,i}\frac{\partial l_{n,i}}{\partial y_{n,i}}+
l_{n+1,i}\frac{\partial l_{n+1,i}}{\partial
y_{n,i}}\,\right]\,,\label{eq:pydot}
\end{eqnarray}
with the derivatives
\begin{eqnarray}
\frac{\partial d_n}{\partial
x_{n,i}}&=&\frac{(-1)^{i+1}\,(d_n^x+x_{n,1}-x_{n,2})}{\sqrt{
\left(\,d_n^x+x_{n,1}-x_{n,2}\right)^2\,+\,
\left(\,d_n^y+y_{n,1}-y_{n,2}\right)^2}}\,,\\ \frac{\partial
l_{n,i}}{\partial x_{n,i}}&=&
\frac{L_x+x_{n,i}-x_{n-1,i}}{\sqrt{h^2+
\left(\,L_{n,i}^x+x_{n,i}-x_{n-1,i}\right)^2\,+\,
\left(\,L_{n,i}^y+y_{n,i}-y_{n-1,i}\right)^2}}\,,
\end{eqnarray}
and the equivalent expressions for ${\partial d_n}/{\partial
y_{n,i}}$ and ${\partial l_{n,i}}/{\partial y_{n,i}}$.

It is useful to compare our model with previous ones:
\begin{enumerate}
\item Refs.~\cite{BCP,biophys} add a three body curvature term
$\sum G_0 \,(\phi_{n+1}+\phi_{n-1}-2\,\phi_n)^2$ with a quite
large constant $G_0=50\,\mathrm{eV/\AA^{-1}}$,  the elastic
constant being $K=1\,\mathrm{eV/\AA^2}$.
 The reason to introduce this term is to guarantee the correct
 helical shape instead of an alternatively possible zig-zag one.
\item Refs.~\cite{Cocco99,Cocco} discard the term above and add
a {\em stacking} term
$E\,\exp\large(-b(r_n+r_{n+1}-2\,R_0)\large)\,(r_n-r_{n-1})^2$,  a
extra degree of freedom, the axial displacement $h$ with an {\em
elastic} energy  $K\,[h_n-h]^2$ instead of the covalent harmonic
bond along the strand.
  The stacking term is proposed to describe the shear force that
  opposes sliding motion of one base over another. The exponential
  attenuation becomes important for large opening at temperatures
  close to denaturation because the decrease of the molecular
  packing. The elastic energy with $h<l_0$ guarantees the helical
  shape. The value of $E=4\,\mathrm{eV/\AA^2}$ is quite larger
  than $K=0.014\,\mathrm{eV/\AA^2}$, bringing about probably the
  dominance of the stacking term over covalent terms.

\item Ref.~\cite{Campa01} maintains the covalent energy term, discards
the curvature term and adds a {\em stacking} term but without the
attenuation factor, as he does not consider high temperatures.
 (See the discussion of the different terms in this reference).
\item Ref.~\cite{Agarwal} uses the harmonic {\em stacking} term on the
radial variables, the curvature term, the covalent bond term and
the vertical elastic term.
\end{enumerate}

 In our work we do not introduce all these terms, because we are
 interested in the effect of the helical shape in itself on
 breather formation, and they are not deduced from first
  physical principles.
  This makes it incomplete,
 as there are,
 certainly, forces that maintain the helical shape, but we
think that the introduction of extra energy terms would mask it.

 We think that an appropriate deduction of the form of the
 stacking energy terms is still to come.

We use the following values of the scaled parameters  $K=0.683$,
$r_0=44.50$, $h=15.13$ and $l_0=31.39$. One time unit of the
scaled time corresponds to $0.198\,ps$ of the physical time.

\section{Energy redistribution, relaxation dynamics and breather formation}

In the following we consider the energy exchange process between
the radial and torsional degrees of freedom when a confined region
of the DNA molecule gets deformed from its equilibrium
configuration. Nowadays, there exist sophisticated experimental
techniques for the selective excitation of DNA in single molecule
experiments (see, e.g.,
\cite{Essevaz},\cite{Clausen}) and during the
last years, mechanical properties of DNA molecules have received a
lot of attention \cite{Marko}-\cite{Clausen}. Several force
measurements  were performed on single DNA molecules to examine,
e.g., the longitudinal extension \cite{Smith} and the twist
elasticity of DNA molecules
\cite{Marko}. Furthermore, the opening
of two-stranded DNA molecules was mechanically forced by pulling
apart the two strands of a DNA double helix \cite{Bockelmann}.

Besides the study of the relaxation process in deformed DNA
molecules, our aim is also to create spatially extended H-bridge
breather solutions, reproducing the oscillating 'bubbles' observed
for the DNA-opening process, which extend over $15-20$ base pairs
\cite{Barbithesis}. To this end we excited initially twenty
consecutive lattice sites in the center of the DNA lattice
assuming
 that the DNA molecule experienced deformations in radial direction. For
the numerical simulation we elongated each of the associated twenty hydrogen
bonds from its equilibrium length by displacing the Morse
oscillators out of their rest positions accordingly.
Provided these elongations are aligned solely along the
equilibrium orientation of the hydrogen bonds no twist deformations
occur (yet).
  Naturally, the distortion
of a hydrogen bond is connected with a deformation of the covalent
bonds of the phosphate backbone in the neighborhood of the
corresponding base pair. Nevertheless, for the starting excitation
pattern, with radial amplitudes in the range of
$d_{\{n_c\}}=(0.1-0.35)$\,\AA, one finds only small deformations
of the covalent bonds. The set $\{n_c\}$ labels the indices of the
twenty excited lattice sites in the central region.
 Compared
to the equilibrium DNA conformation the deformed one possesses an
amount of potential energy increased by the deformation energy
(also referred to as excitation energy). Furthermore, the energy
contained in the elongated hydrogen bonds, $U_h$,
is exceedingly larger than those contained in the deformed
covalent bonds, $U_c$,
and the ratio is typically of the order of $U_c/U_h\lesssim 0.15$.

\begin{figure}
  \begin{center}
    \includegraphics[height=0.4\textheight]{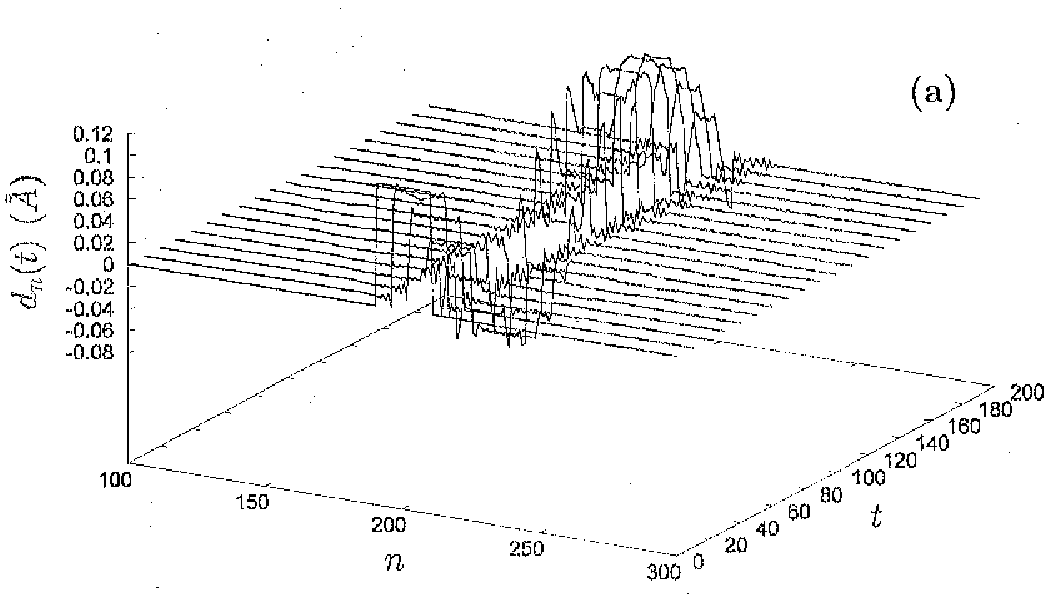}
    \includegraphics[height=0.4\textheight]{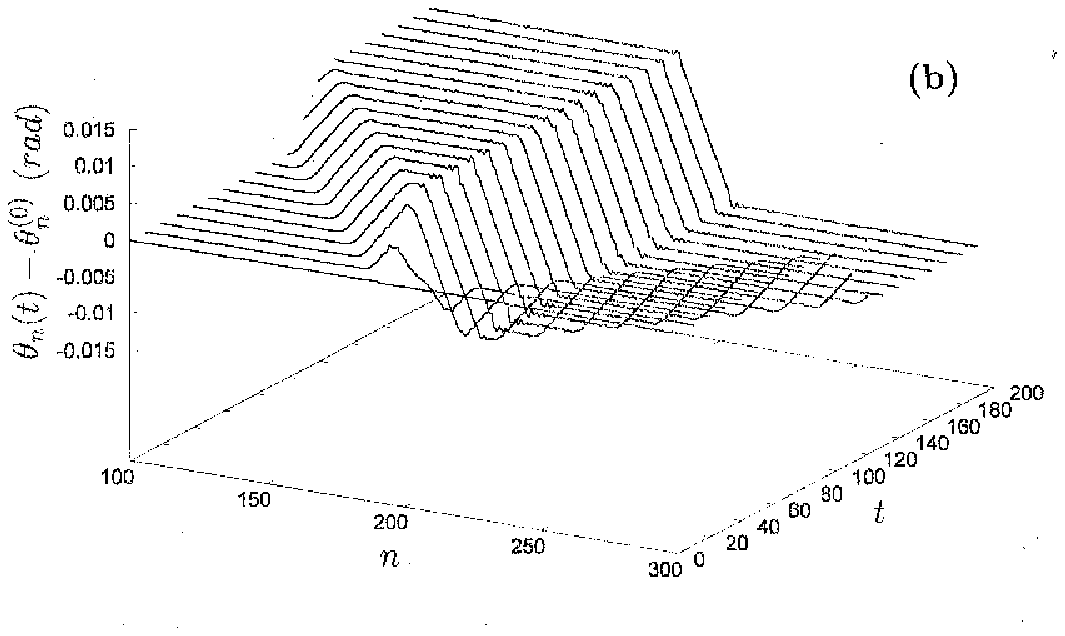}

\end{center}
  \caption{Relaxation dynamics:  For a proper
illustration of the extension of the radial breather and the
forming kink-like angular structure an inner segment of the DNA
lattice consisting of $200$ sites is shown. Initial conditions:
$d_{\{n_c\}}=0.112$\,\AA. (a) Radial breather.(b) Angular
deformation pattern. Development of the kink-like structure. }
  \label{fig:fig2}
  \end{figure}

We integrated the set of coupled equations
(\ref{eq:xdot})-(\ref{eq:pydot}) with a fourth-order Runge-Kutta
method while the accuracy of the computation was checked through
monitoring the conservation of the total energy. For the
simulation the DNA lattice consists of $400$ sites and open
boundary conditions were imposed. (We remark that, provided the
system size is sufficiently large, our findings are insensitive to
further enlargement of the system.) First of all, we report on the
results obtained for the regular equilibrium configuration of the
DNA molecule for which the bases of a pair are rotated around the
axis by the same angle
$\theta_{n,1}=\theta_{n,2}+\pi=\theta_n^{(0)}=n\theta_0$.
Consequently, the equilibrium twist angle between two consecutive
base pairs is $\theta_0$ and all bases possess an equal radial
distance $d_0/2$ from the central axis. In Fig.~\ref{fig:fig2}--a
we depict the spatio-temporal evolution of the distance, $d_n(t)$,
between two bases of a base pair, which measures the variation of
the length of the corresponding hydrogen bond. Initially, the
excitation energy is divided evenly between the twenty lattice
sites so that the initial distance profile, $d_n(0)$, is
rectangularly shaped. In the current regular case, the two bases
of a base pair move symmetrically with respect to the central axis
in radial direction due to the asymmetric choice of initial
conditions, $(x_{n,1}(0),y_{n,1}(0))=-(x_{n,2}(0),y_{n,2}(0))$.
Thus, the associated displacement coordinates perform out-of-phase
oscillations $(x_{n,1}(t),y_{n,1}(t))=-(x_{n,2}(t),y_{n,2}(t))$.
In addition, the two bases of a pair possess symmetric angular
dynamical displacements,
$\theta_{n,1}(t)=\theta_{n,2}(t)+\pi=\theta_n(t)$, and
 the line connecting the two (elongated) bases in the base
plane always intersects the central axis. The distance variable
$d_n(t)$ is hereafter also referred to as the radial variable
because $d_n(t)/2$ represents actually the local helix radius. The
corresponding angular displacement pattern,
$\theta_n(t)-\theta_n^{(0)}$, is shown in Fig.~\ref{fig:fig2}--b.

As becomes evident from Fig.~\ref{fig:fig2}--a the vast majority
of the excitation energy remains in the initially excited central
region and for the radial motion a multi-site breather develops
for which the stretching of a base pair distance is larger than
the compression characteristic for the evolution in a Morse
potential (see also \cite{BCP}). Moreover, the resulting spatially
extended radial breather involves all of the initially excited
twenty oscillators evolving with practically equal amplitudes. To
either side of the central region the amplitude pattern abruptly
goes to zero giving the multi-site breather an almost rectangular
spatial profile nearly matching the initial profile. The
maintenance of the broad localized radial shape has to be
distinguished from the dynamical behavior reported in
\cite{biophys} where, in the context of the BCP model, for similar
initial conditions in the radial stretchings in a fairly broad
region a
 merging of the radial components  is observed in the course of time which results eventually
in
 a narrow radial pattern localized at a single site only.

For the associated angle deformations
$\theta_n(t)-\theta_n^{(0)}$, emerging from overall initial zero
amplitudes, an asymmetric pattern generates rather rapidly in the
initially excited region. It holds that the angles from the left
(right) of the two central lattice sites (base pairs) decrease
(increase) steadily and, the further apart a base pair is from the
central ones, the stronger its angular displacement. Eventually, a
local kink-like pattern in the angular lattice
$\theta_n(t)-\theta_n^{(0)}$ is created. The horizontal plateaus
of the kink-pattern extend continuously in either direction away
from the central base pairs so that in the course of time more and
more base pairs become subject to angle deformations leading to a
progressive untwisting of the central part of the helix
\cite{Moroz}. This untwisting of the helicoidal helix structure
results from the coupling between the radial and the torsional
degrees of freedom due to geometrical constraints and is typical
for the DNA opening dynamics \cite{BCP}. In contrast to the
periodically oscillating pattern of the radial variable $d_n(t)$,
corresponding to alternate stretchings and compressions of the
hydrogen bonds, the torsional deformations
$\theta_n(t)-\theta_n^{(0)}$ adjust to a static deformation
pattern. For a direct comparison of the extension of the radial to
the angular helix deformations, the angular component, expressed
in radians, has to be multiplied by $r_0\thickapprox 10$\,\AA\, to
estimate the associated length scale. We conclude that the degree
of the deformations in the angular direction are comparable with
those in the radial direction. This has to be distinguished from
the observations made for the evolution of narrow radial breathers
for which the dynamics of the opening process is dominated by the
displacements of the bases in radial direction whose amplitudes
are typically larger than those of the angular displacements by at
least one order of magnitude \cite{BCP}.

For further illustration of the energy sharing phenomena
accompanying the breather formation process, we
monitored the time-evolution of the energy contained in the
initially excited $20$ base pairs, which is determined by
\begin{equation}
E_{center}(t)=\sum_{\{n_c\}}\,\left\{\,\left[\,
\sum_{i=1,2}\,E_{kin}(p_{n,i}^{(x)},p_{n,i}^{(y)})
+U_b(l_{n,\,i})\,\right]+U_h(d_{n}) \,\right\}\label{eq:Ecenter}.
\end{equation}
The temporal behavior of $E_{center}(t)$ is represented by the
graph with label $(1)$ in Fig.~\ref{fig:fig6}. We observe that in
an early phase, lasting for approximately forty time units ($\sim
8\,ps$), a small energy loss takes place and  around $1\%$ of the
excitation energy flows from the initially excited central region
into wider parts of the remainder of the DNA lattice. Afterwards
the amplitude of the energy $E_{center}(t)$ fulfills slow
oscillations around a mean value. (That this mean value of the
energy $E_{center}(t)$ is weakly descending as time progresses
indicates further small energy emission from the central region
into the rest of the DNA lattice.) The energy emitted from the
central region is then mainly redistributed into the angular
deformation kink that travels through the DNA lattice
 in agreement with the formation of the
 angular  pattern represented in Fig.~\ref{fig:fig2}--b.
 Nonetheless,
 a tiny part of the emission energy is radiated also into
the hydrogen bonds outside the central region where it is
converted in radial phonons moving uniformly towards the ends of
the lattice as just about visible in Fig.~\ref{fig:fig2}--a.

\begin{figure}
  \begin{center}
    \includegraphics[width=\singlefig]{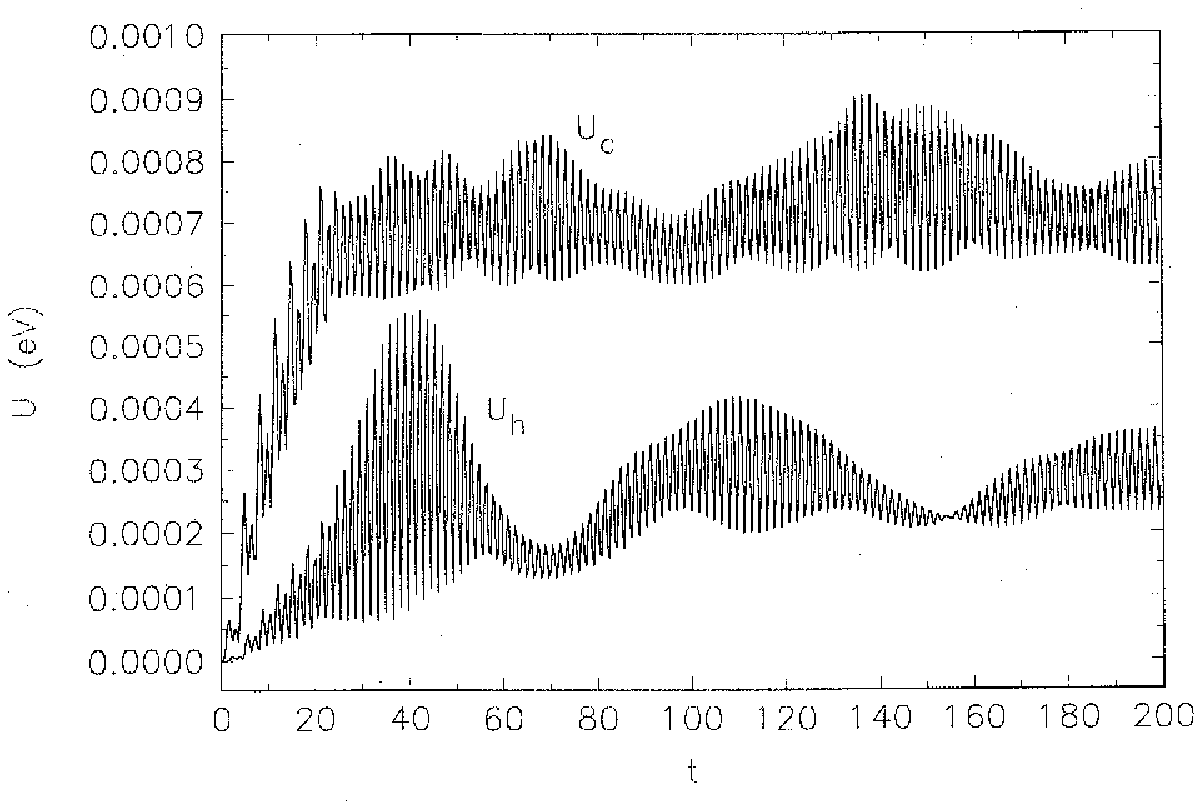}
  \end{center}
  \caption{Relaxation dynamics: The distribution of
potential energy between the hydrogen bonds and the covalent bonds
outside the central region of the DNA lattice.}
  \label{fig:fig3}
\end{figure}

The Fig.~\ref{fig:fig3} shows the distribution of the energy,
between the hydrogen bonds and the covalent bonds, which has been
deposited from the central region into the rest of the DNA
lattice. Plotted are the time-evolution of the potential energy
contained in the displaced hydrogen bonds
\begin{equation}
U_{h}^{c}(t)=\sum_{\tilde{n}}\,(\exp(-d_{n}(t)) -1)^2\,,
\end{equation}
 and the potential energy content of the covalent bonds
\begin{equation}
U_{c}^{c}(t)=K\sum_{\tilde{n},\,i=1,2}\,l_{n,i}^2(t)\,,
\end{equation}
respectively, where $\tilde{n}\in [1,N]\setminus \{n_c\}$.
After the initial phase of energy absorption a quasi-equilibrium
regime is reached and the partial potential energies  oscillate
around mean values of $0.0007\,eV$ and $0.0003\,eV$ for the
deformations of the covalent bonds and  the hydrogen bonds,
respectively.

\begin{figure}
  \begin{center}
    \includegraphics[width=\singlefig]{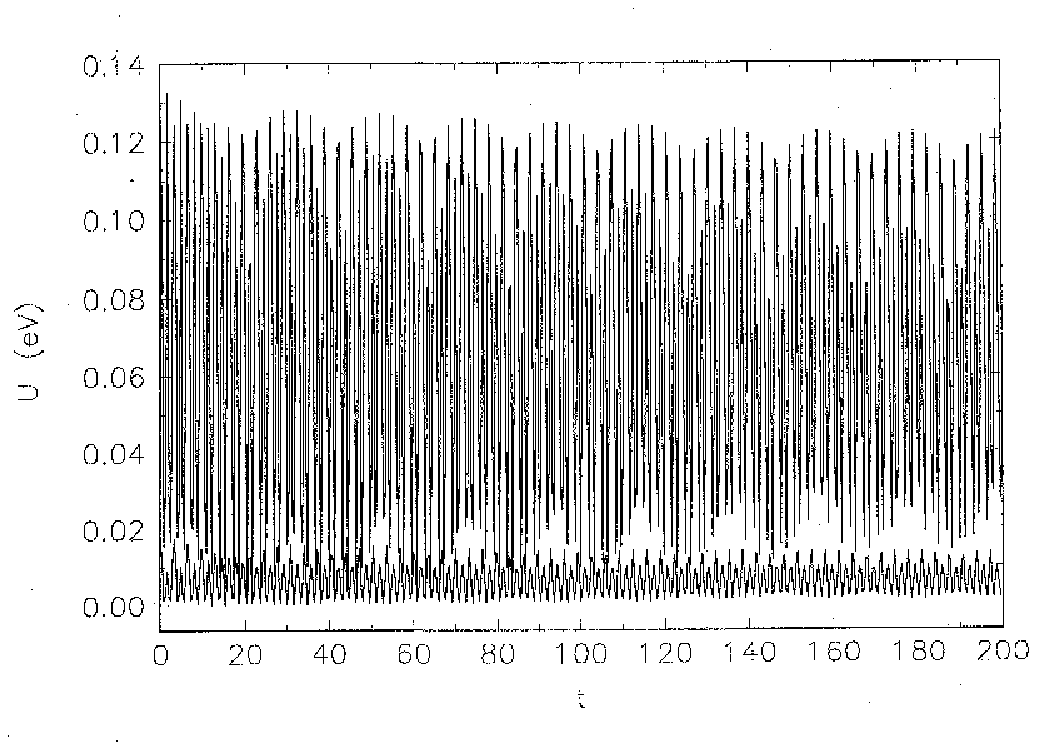}
  \end{center}
  \caption{Relaxation dynamics:  The energy
sharing between the deformed covalent bonds (upper graph) and
hydrogen bonds (lower graph) within the initially excited region.}
  \label{fig:fig4}
\end{figure}

Complementary, the energy sharing between the deformed covalent
and hydrogen bonds within the initially excited region is
illustrated in Fig.~\ref{fig:fig4}. During a transient phase of
internal energy redistribution the average of the covalent bond
energy grows slightly on expense of the hydrogen bond energy.
However, the energy migration is not very pronounced. Eventually,
the time evolution of each of the two partial potential energies
is characterized by periodic oscillations around a (constant) mean
value reflecting the attainment of a steady equilibrium regime in
the central lattice region. Note that the initial drastic
difference in the energy contained in the stretched hydrogen
bonds, $U_h(0)=0.124\,eV$, and the covalent bonds,
$U_c(0)=0.018\,eV$, is retained throughout the time-evolution of
the respective average partial potential energy.

In conclusion, we found that out of an initial non-equilibrium
situation, for which the hydrogen bonds in a fairly broad but
confined region of the central part of the DNA lattice have been
stretched away from their rest lengths, (weak) energy
redistribution processes set in, such that small amounts of energy
migrate into the remainder of the DNA lattice. During the
relaxation process towards an equilibrium of energy balance
between the radial and torsional components the angular
displacement variables adopt a static kink-like structure.
Correspondingly, due to the torsional deformations induced by the
opening of the base pairs a local unwinding of the helix develops.
Significantly, almost the entire initial excitation energy stays
localized in the hydrogen bonds of the central region and a {\it
multi-site breather} with nearly rectangular profile develops in
the radial displacement variables.
\begin{figure}[h]
  \begin{center}
    \includegraphics[width=\doublefig]{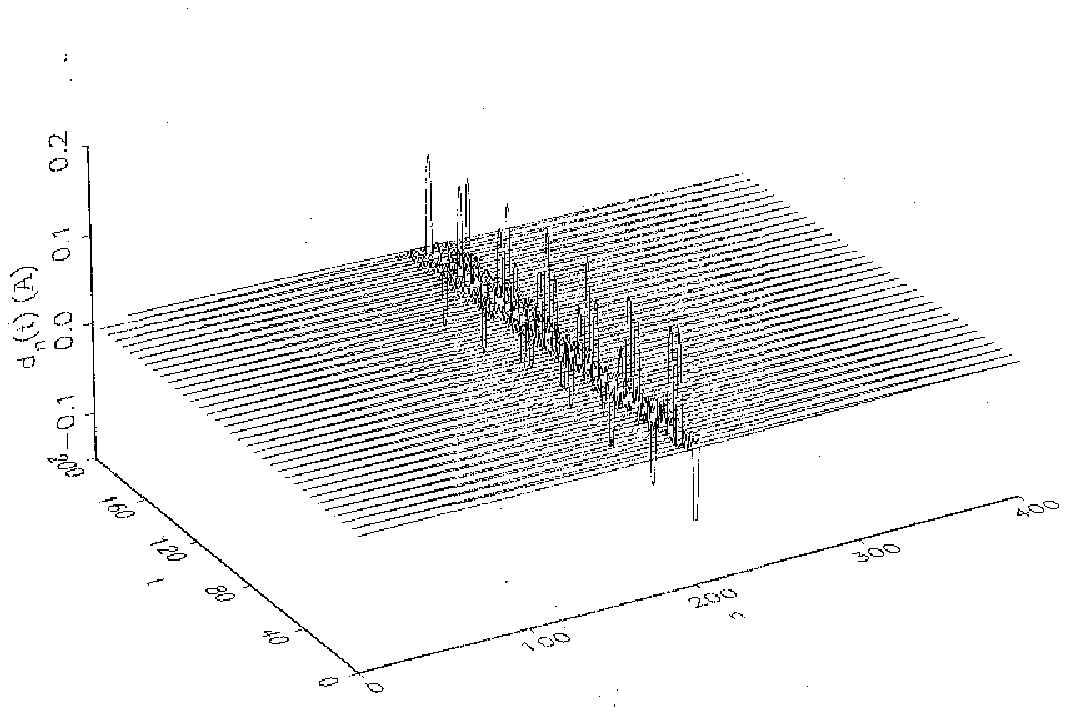}
  \end{center}
  \caption{H-bond breather formation for asymmetric initial conditions.
 Three consecutive
sites on one strand possess initially diminished distance to the
helix axis. }
  \label{fig:fig5}
\end{figure}

Besides the initial pattern of broad radial inflation we imposed initially  also
radial compressions in a fairly broad region of the DNA lattice giving also rise to
  multi--site breathers corresponding to alternating compressions and stretchings
of the H-bonds.

Finally, regarding {\it asymmetric} initial conditions we mimicked e.g. the impact of a local
 'pushing' force in cause of which  some bases are brought closer to the helix axis.
 In Fig.~\ref{fig:fig5} we display the spatio-temporal evolution of the radial deformations
of the H-bonds when initially three nucleotides at only one strand
were displaced from their equilibrium positions. Apparently, even
for such asymmetric situations  H-bond breathers result
characterized by deformations of the hydrogen bonds which are no
longer symmetric with respect to the central axis.
  The conclusion
is that the multibreathers are not destroyed by interaction with
the modes of the centers of mass . Other numbers of excited base
pairs bring about similar results.
 \section{The relaxation concept and breather construction}

We exploit now the findings of the preceding section for the
construction of multi-site breathers in DNA. For this goal we
initialize the DNA lattice as described above, that is twenty
Morse oscillators in the central part of the DNA lattice are
elongated from their equilibrium lengths. Subsequently, we let the
 DNA lattice dynamics  relax towards an equilibrium regime during
 a time interval of $100$ time units. We recall that for such
 times the energy redistribution between the
 hydrogen bonds and the covalent bonds and the
 central region and the remainder of the DNA lattice, respectively
 has proceeded
 so far, that virtually negligible alterations of the partial energy
 contents occur for later times. Notably, a spatially extended
 radial breather forms on such a time scale.
 Furthermore, due the geometrical
 constraint of the helicoidal structure a local untwisting
 of the double helix arises. The corresponding kink-like pattern in the
 angle variables
encompasses almost the whole lattice, apart from the small parts
of the lattice near the boundaries which have not yet experienced
angle deformations (cf. Fig.~\ref{fig:fig2}--b).

\begin{figure}
  \begin{center}
    \includegraphics[width=\singlefig]{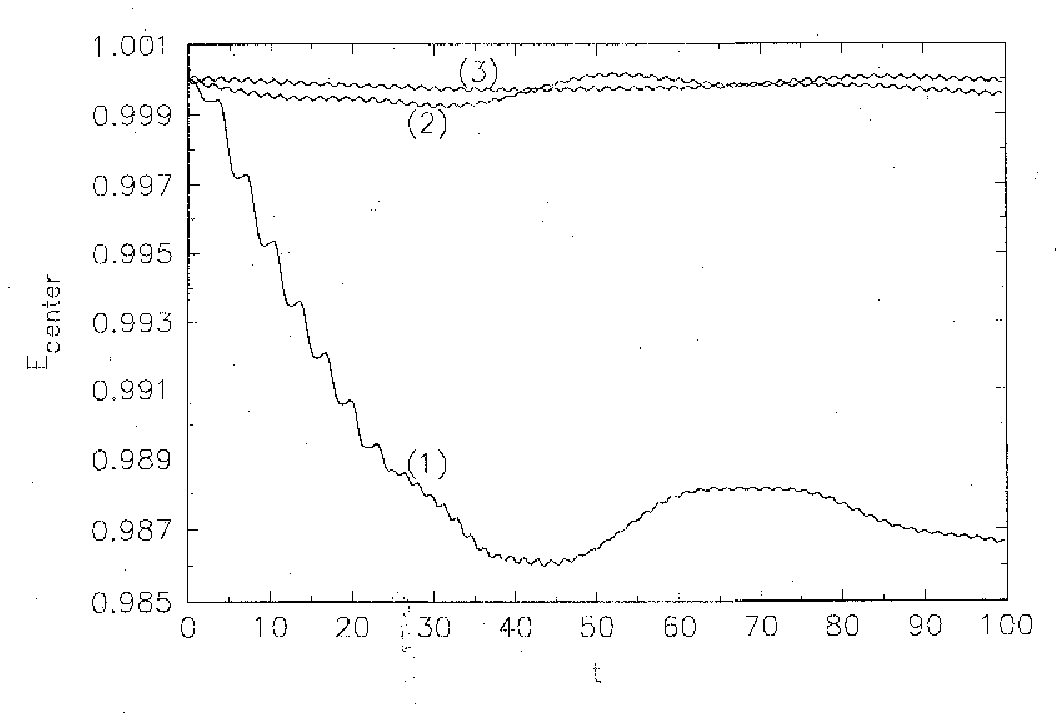}
  \end{center}
  \caption{Breather construction: The temporal evolution of the energy
contained in the central part of the lattice for the first, the
second
 and the third step of the iteration procedure, respectively,
 as indicated in the plot. The energies are normalized in units of
the initial energy of the first iteration $E_{center}(0)$.}
  \label{fig:fig6}
\end{figure}

\begin{figure}
  \begin{center}
    \includegraphics[width=\doublefig]{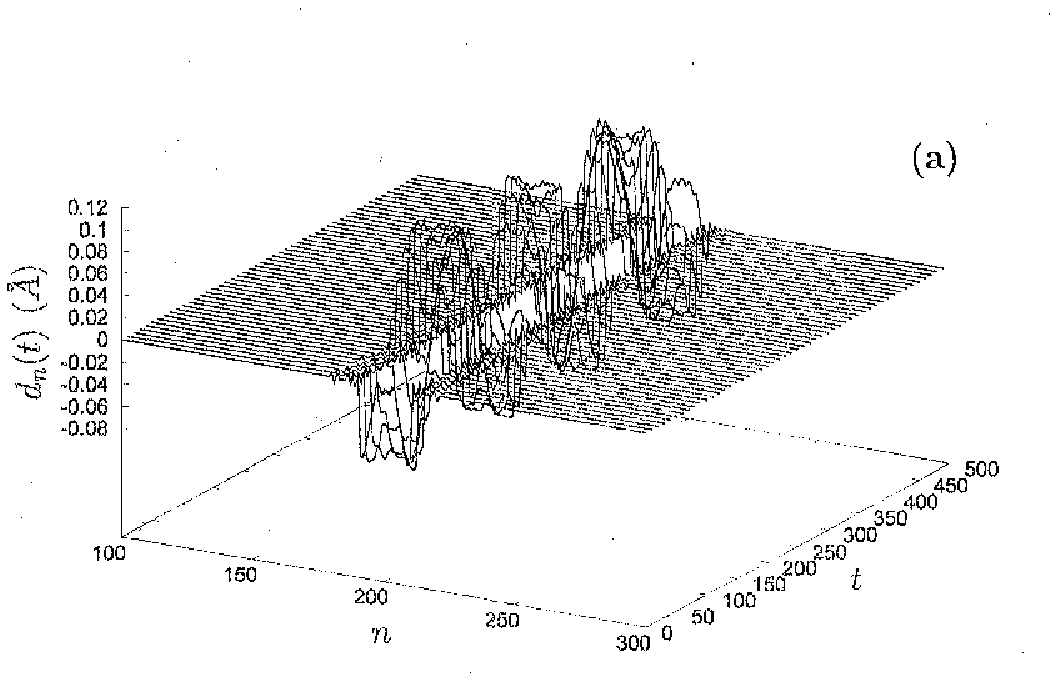}
    \includegraphics[width=\doublefig]{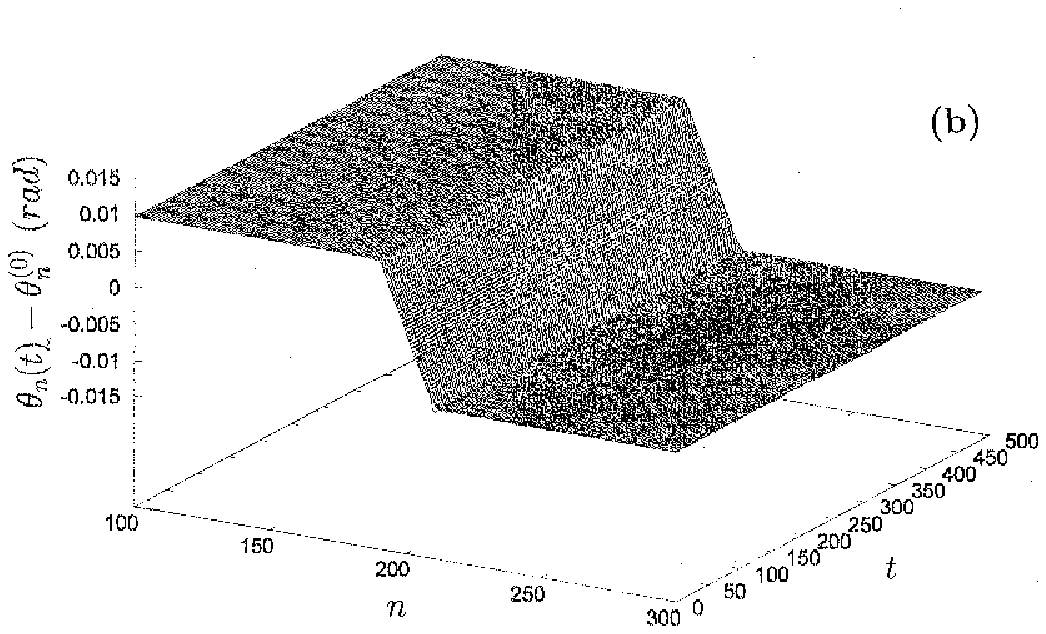}
  \end{center}
  \caption{Breather construction: The final radial
breather solution obtained after three iteration steps of the
relaxation concept.}
  \label{fig:fig7}
\end{figure}

We start now a iteration scheme, for which the quasi-equilibrium
solution resulting at the end of the simulation time, viz.
the displacements
$(x_{n,i}(100)$, $y_{n,i}(100))$, is taken as the new initial
condition for the re-iterated DNA lattice dynamics. In order to
attain an equilibrium regime of the DNA lattice dynamics the
moving radial phonons, carrying the excess energy emitted from the
central region, have to be eliminated. In the primary iteration
step the radial phonons  transport energy merely of the order of
$10^{-4}\,eV$. With each further iteration cycle the amount of
radial phonon energy, still ejected  from the central lattice region,
further diminishes. Nevertheless, to accomplish faster relaxation
we impose absorbing boundary conditions to the DNA lattice so that
the radial phonons
 are removed, once they arrive at the ends of the
lattice. In this iterative manner the energy redistribution from
the central region to the remainder of the DNA lattice is more and
more reduced and the dynamics relaxes towards a solution of
improved energy balance compared to the previous step. To
illustrate the latter fact, we display in Fig.~\ref{fig:fig6} the
temporal evolution of the energy, $E_{center}(t)$, contained in
the central part of the lattice which bears the radial breather.
Apparently, convergence of the iteration procedure is achieved
already after three iteration steps. In Fig.~\ref{fig:fig7} we
depict the resulting radial breather solution. This breather
oscillates with a period of $=0.7\,ps$ which is a realistic value
for the time scale of experimentally observed vibrational modes
occurring prior to the opening process in DNA melting
\cite{Prohofsky}. More importantly, the broad localized radial
excitation pattern comprises twenty base pairs and the maximal
elongation of a base pair radius from rest length is about
$0.122$\,\AA\,. Thus, we are able to create long-lived spatially
extended radial breathers of fairly large amplitude in DNA
representing the oscillating 'bubbles' as the precursor to thermal
denaturation. Finally, we remark that, depending on the choice of
the width of the initial excitation pattern, with our
relaxation-method radial DNA breathers of varying extension can be
generated. These breathers range up from narrow ones (effectively
one-site) up to very broad breathers involving up to even fifty
lattice sites. As the choice of the initial excitation pattern is
further concerned, we tried, besides the radial rectangular
profiles, also other profiles such as bell-shaped ones given by
$d_{\{n_c\}}(0)=A_0/\cosh[\beta\,(n-n_0)]$. $A_0$, $\beta$ and
$n_0$ are constants fixing the amplitude, the spatial extension
and the center of the starting profile, respectively. In the
outcome of the relaxation dynamics we observed bell-shaped radial
breathers nearly preserving the initial profile analogous to the
results reported earlier in this manuscript for the broad
breathers. The same holds when additionally to the radial
displacements also initial  kink-like twist deformations are
considered.

For a more elaborate study of  nonlinear vibrational modes in DNA
the effects of external influences have to be included. Real DNA
molecules exhibit random structural imperfections of their double
helix caused, e.g., by the deforming impact of the chemical
surroundings when DNA  gets buffeted by water molecules. Another
source for structural irregularities originates from the random
base sequence of the genetic code. In addition, the varying
hydrophobic potential of the base pair interactions depending on
the ambient aqueous solvent may leave the helix structure in an
irregularly distorted shape. Accordingly,  for an improvement of
our model, irregularity effects can be mimicked  by taking into
account structural disorder so that the resulting irregular
helical DNA matrix deviates from the perfectly regular helix
structure. (The formerly discussed ordered, regular structure
arises, for example, for synthetically produced DNA molecules
consisting of a single type of base pairs, for instance,
poly(G)-poly(C) DNA polymers, surrounded by vacuum.)

To be precise, we consider randomly distributed equilibrium
coordinates, $\tilde{x}_{n,i}^{(0)}$ and $\tilde{y}_{n,i}^{(0)}$,
of the bases with $\tilde{x}_{n,i}^{(0)}-x_{n,i}^{(0)}\in[-\Delta
x_{n,i} , \Delta x_{n,i}]$ and
$\tilde{y}_{n,i}^{(0)}-y_{n,i}^{(0)}\in[-\Delta y_{n,i} , \Delta
y_{n,i}]$. The intervals of the deviations $\Delta x$ and $\Delta
y$ range up to $5\%$ of the respective rest value $x_{n,i}^{(0)}$
and $y_{n,i}^{(0)}$ of the corresponding regular structure.

\begin{figure}[h]
  \begin{center}
    \includegraphics[width=\doublefig]{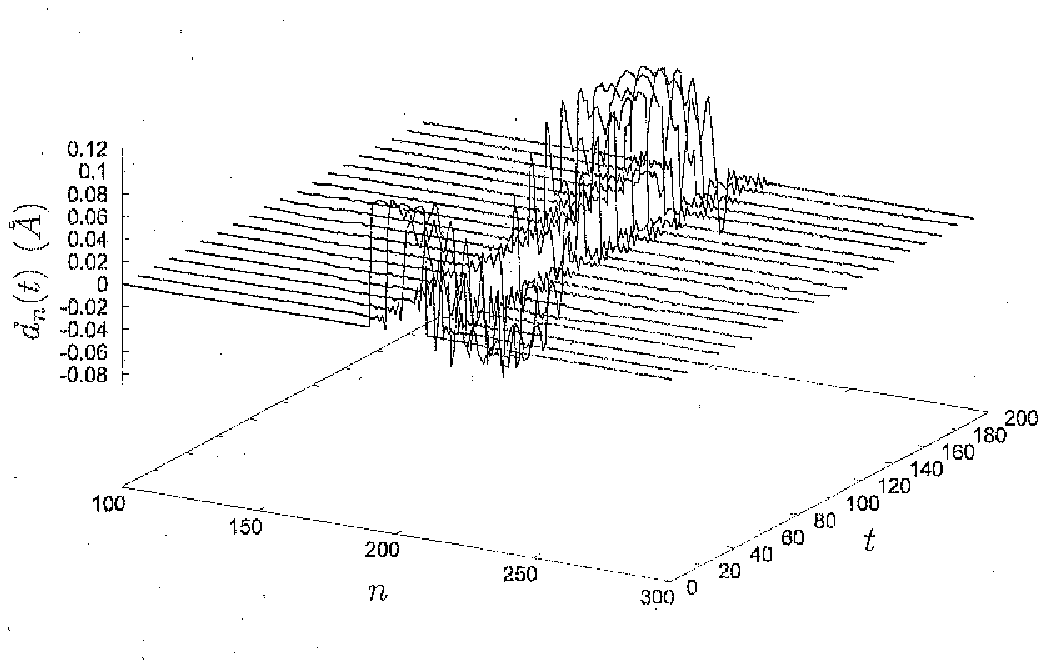}
  \end{center}
  \caption{Breather construction: The persisting radial breather
 for a disordered helicoidal equilibrium structure (see text).
The initial values for the irregular distance displacements in the
central region, $\tilde{d}_{\{n_c\}}$, are randomly distributed
according to $\tilde{d}_{\{n_c\}}-d_{\{n_c\}}\in[-\Delta d ,
\Delta d]$. The interval of the deviations $\Delta d$ ranges up to
$5\%$ of the respective  value $d_{\{n_c\}}$ of the regular case
depicted in Fig.~\ref{fig:fig2}. }
  \label{fig:fig8}
\end{figure}

In Fig.~\ref{fig:fig8} we show a representative case for the DNA
lattice dynamics for one realization of structural disorder. In
addition to the disordered helicoidal structure also random
amplitudes of the initial values for the distance displacements,
$d_{\{n\}}(0)$, in the central region have been chosen.
Apparently, the H-bridge breather is robust sustaining the
influence of disorder and virtually resembles the behavior of the
ordered case (compare Fig.~\ref{fig:fig2}--a).

\begin{figure}
  \begin{center}
    \includegraphics[width=\doublefig]{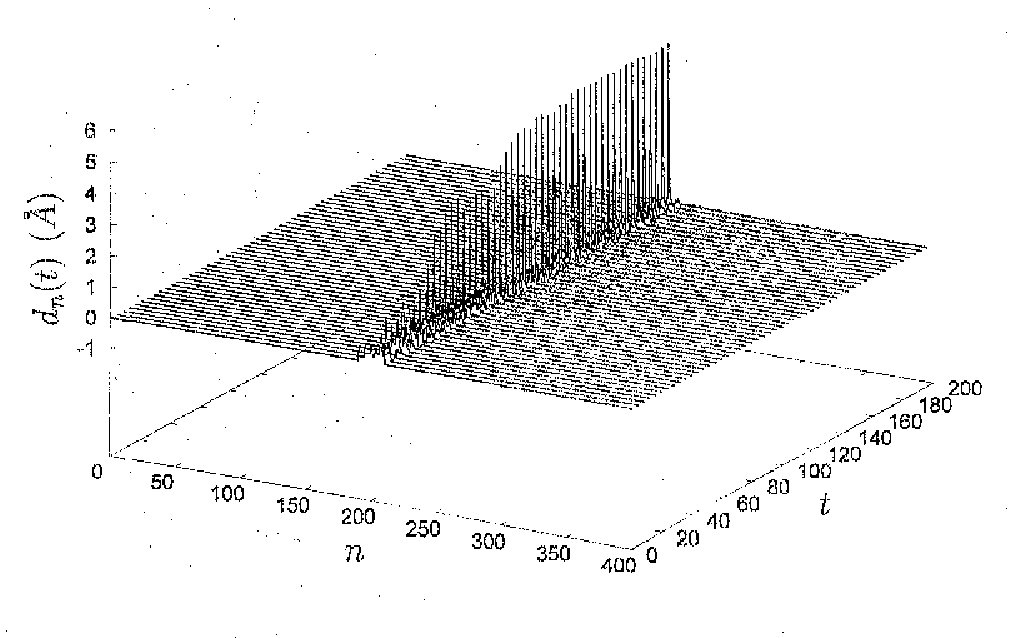}
  \end{center}
  \caption{Concentration of the excitation energy
on a single site of the DNA lattice and successive bond breaking.
Initial conditions corresponding $d_{\{n_c\}}=0.75$\,AA
corresponding to an amount of excitation energy of $0.646\,eV$.}
  \label{fig:fig9}
\end{figure}

On the other hand, when the amount of excitation energy placed in
the central lattice sites,  exceeds a critical value,
$E_{n_c}\gtrsim 0.6\,eV$, or equivalently when the hydrogen bonds
get elongated too far, we do not longer observe the relaxation
towards a stable multi-site radial breather as in the previous
(lower-energy) cases. Instead directed flow of energy into one of
the initially excited lattice sites takes place related with the
simultaneous energy depletion of the remaining lattice sites. In
this manner, the transferred energy is accumulated in the
corresponding hydrogen bond of the energy-gaining site with the
result that the radial elongation continually grows. Eventually,
the corresponding hydrogen bridge is stretched so far that it
breaks up. This energy concentration onto a single site and the
succeeding bond breaking process is illustrated in
Fig.~\ref{fig:fig9}.

 Conversely, even for as tiny radial displacement amplitudes
as $d_{\{n\}}(0)\sim 10^{-5}$\,\AA, corresponding to a lattice
excitation energy of the order of $10^{-7}\,eV$, the energy stays
localized at the twenty initially lattice sites and a radial
multi-site breather is formed. This demonstrates impressively that
DNA possesses very efficient energy storing abilities. It should
be mentioned, that usually, in studies of nonlinear oscillator
networks one observes that for undercritical amplitudes, being
tantamount to a low degree of inherent nonlinearity, an initially
localized structure decays rather rapidly and the excitation
energy gets uniformly distributed over the whole lattice
\cite{Flach}.

It should be noted that Campa \cite{Campa01} also shows the
existence of broad bubbles that move along the molecule. The
differences between his model and ours are commented in the
previous section. The results are quite different. The Campa
bubbles have no oscillation and a bell profile, while our
multibreathers oscillate and have a quasi rectangular profile,
involving a larger number of base pairs in a similar way. The
oscillation is a major weakness of our bubbles as they can only be
considered as precursors of transcription or denaturation. They
have some advantage, however, which are their quasi homogenous
amplitude and that they are more generic, being the consequence of
the helical shape, and not relying on evolving in the flat part of
the Morse potential.

\section{Linear modes}
\label{sec:linear}
\newcommand{\ii}{\mathrm{i}}
\newcommand{\bp}{{\em bp }}
\newcommand{\cm}{{\em cm }}

In this section we consider the linear modes of the system and their influence on
the behavior and stability of the system. The linear equations corresponding
to Eqs.~(\ref{eq:xdot}--\ref{eq:pydot}) for the homogeneous and symmetric system are given by:
\begin{eqnarray}
&\mbox{}&\ddot{x}_{n,i}=(-1)^i\,2\left(d_n^x\,\Delta x_n +d_n^y\, \Delta y\right)\frac{ d_n^x}{d_0^2}\label{eq:linx}\\
&\mbox{}&-2\,K(\frac{L_{n,i}^x}{l_0^2}[L_{n,i}^x\delta x_{n,i} + L_{n,i}^y\delta y_{n,i}]
-\frac{L_{n+1,i}^x}{l_0^2}[L_{n+1,i}^x\delta x_{n+1,i} + L_{n+1,i}^y\delta y_{n+1,i}]\,)\,,
\nonumber\\
&\mbox{}&\ddot{y}_{n,i}=(-1)^i\,2\left(d_n^x\,\Delta x_n +d_n^y\, \Delta y\right)\frac{ d_n^y}{d_0^2}\,,
\label{eq:liny}\\
&\mbox{}&-2\,K(\frac{L_{n,i}^y}{l_0^2}[L_{n,i}^x\delta x_{n,i} + L_{n,i}^y\delta y_{n,i}]
-\frac{L_{n+1,i}^y}{l_0^2}[L_{n+1,i}^x\delta x_{n+1,i} + L_{n+1,i}^y\delta y_{n+1,i}]\,)
\nonumber
\end{eqnarray}
where $\Delta x_n=x_{n,1}-x_{n,2}$, $\Delta y_n=y_{n,1}-y_{n,2}$,
$\delta x_n=x_{n,i}-x_{n-1,i}$ and $\delta y_n=y_{n,i}-y_{n-1,i}$.
In the symmetric equilibrium state it holds that $L_{n,2}^x=-L_{n,1}^x=L_0\cos(\phi_n^{(0)})$,
$L_{n,2}^y=-L_{n,1}^y=L_0\sin(\phi_n^{(0)})$, $d_{n,x}=d_0\,\cos(\theta_n^{(0)})$, $d_{n,y}=d_0\,\sin{\theta_n^{(0)}}$, with $L_0=((L_n^x)^2+(L_n^y)^2)^{1/2}=d_0\sin(\theta_0/2)$, being the equilibrium horizontal distance between nucleotides
along one strand; $\phi_n^{(0)}$ being the angular cylindrical
coordinate of the vector that goes from nucleotide $n-1$ to $n$
along  strand $1$  at equilibrium, i.e., $\phi_n^{(0)}=\theta_n^{(0)}-\theta_0/2 + \pi/2$; and $\theta_n^{(0)}=n\,\theta_0$, as defined in Eq.~\ref{eq:thetan}, at equilibrium.

The natural coordinate system for each base pair is given by the projections on the unit vectors  $\mathbf{e}_r=(\cos(\theta_n^{(0)}),\sin(\theta_n^{(0)}),0)$ and $\mathbf{e}_\theta=(-\sin(\theta_n^{(0)}),\cos(\theta_n^{(0)}),0)$, and the natural variables are the relative displacements of the base pairs $\Delta\mathbf{r}_n=(\Delta x_n,\Delta y_n,0)= (x_{n,1}-x_{n,2},y_{n,1}-y_{n,2},0)$, and the positions of their centers of mass in the horizontal plane: $\overline{\mathbf{r}}_n=(\overline{x}_n,\overline{y}_n,0)= ((x_{n,1}+x_{n,2})/2,(y_{n,1}+y_{n,2})/2,0)$. Therefore, the natural coordinates are:
\begin{eqnarray}
r_n&=&\Delta\mathbf{r}_n\cdot \mathbf{e}_r = \cos(\theta_n^{(0)})\Delta x_n +\sin(\theta_n^{(0)})\Delta y_n\,,\nonumber\\
\alpha_n&=&\frac{\Delta\mathbf{r}_n\cdot \mathbf{e}_\theta}{d_0} =\frac{1}{d_0}\left( -\sin(\theta_n^{(0)})\Delta x_n +\cos(\theta_n^{(0)})\Delta y_n\right)\,,\nonumber\\
\overline{r}_n&=&\overline{\mathbf{r}}_n\cdot \mathbf{e}_r = \cos(\theta_n^{(0)})\overline{x}_n +\sin(\theta_n^{(0)})\overline{y}_n\,,\nonumber\\
\overline{s}_n&=&\overline{\mathbf{r}}_n\cdot \mathbf{e}_\theta = -\sin(\theta_n^{(0)})\overline{x}_n +\cos(\theta_n^{(0)})\overline{y}_n \,.
\label{eq:newvar}
\end{eqnarray}
To first order in the perturbations $\{x_{n,i}\}$,$\{y_{n,i}\}$, $r_n\simeq d_n$ represent the stretchings of the base pairs; $\alpha_n \simeq \arctan[(d_n^y+\Delta y_n)/(d_n^x+\Delta x_n)]-\theta_n^{(0)}(\mathrm{mod} \,\pi)$, the relative twist angle with respect to the equilibrium position. $\overline{r}_n$ and $\overline{s}_n$ are the displacements of the base pairs centers of mass in the directions of the bonds or orthogonal to them respectively.
 In terms of these variables, the linear dynamical equations (\ref{eq:linx}--\ref{eq:liny}) can be written as:
\begin{eqnarray}
\ddot{\overline{r}}_n&=&-\mu\,\sin^2(\frac{\theta_0}{2})
\,(2\,\overline{r}_n+\overline{r}_{n+1}+\overline{r}_{n-1})-\mu\,\sin(\frac{\theta_0}{2})
\cos(\frac{\theta_0}{2})\,(\overline{s}_{n+1}-\overline{s}_{n-1})\,,
 \nonumber\\
\ddot{\overline{s}}_n&=&\mu\,\sin(\frac{\theta_0}{2})
\cos(\frac{\theta_0}{2})\,
(\overline{r}_{n+1}-\overline{r}_{n-1})-
\mu\,\cos^2(\frac{\theta_0}{2})\,(2\,\overline{s}_n-\overline{s}_{n+1}-\overline{s}_{n-1})
 \,,\nonumber\\
\ddot{r}_n&=&-\omega_0^2\, r_n\nonumber\\&-&\mu\,\sin^2(\frac{\theta_0}{2})
\,(2\,r_n+r_{n+1}+r_{n-1})- \mu\,\sin(\frac{\theta_0}{2})
\cos(\frac{\theta_0}{2})\,d_0\,(\alpha_{n+1}-\alpha_{n-1})\,,
 \nonumber\\
d_0\,\ddot{\alpha}_n&=&\mu\,\sin(\frac{\theta_0}{2})
\cos(\frac{\theta_0}{2})\,
(r_{n+1}-r_{n-1})
-\mu\,\cos^2(\frac{\theta_0}{2})\,d_0\,(2\,\alpha_n-\alpha_{n+1}-\alpha_{n-1})
\,,\nonumber \\
 \label{eq:lincmbp}
\end{eqnarray}
with $\mu=2\,K\,(L_0/l_0)^2$.

The first two equations describe the motion of the helix
as a whole, their corresponding linear modes frequencies will be referred to as the \cm (center of mass)  branch. The last two equations represent the dynamics of
the base pairs and  lead to two branches, referred to as
the \bp (base pair) acoustic and optical branches. Note that the \cm equations are
independent of the \bp equations and so are their modes.

 The frequencies of the \cm  branch are given by:
\begin{equation}
\omega^2=2\mu\,[\,1-cos(\theta_0)cos(q)] \label{eq:branchcm}\, ,
\end{equation}
and a branch of zeros, which represents the possible motions of the helix to other equilibrium positions.

The \bp branches are:
\begin{eqnarray}
\omega^2_\pm&=&\frac{1}{2}\left( \omega_0^2 + 2\,\mu \,[1-\cos(\theta_0)\cos(q)]\right)\nonumber\\
&\pm&\frac{1}{2} \sqrt{ \left( \omega_0^2 + 2\,\mu \,[1-\cos(\theta_0)\cos(q)]\right)^2-
8\,\mu\,\omega_0^2\cos^2(\theta_0/2)\sin^2(q/2)}\,.\quad
\end{eqnarray}
The sign $+$ and $-$ correspond to the \bp optical and acoustic branches respectively.

\begin{figure}[h]
  \begin{center}
    \includegraphics[width=\doublefig]{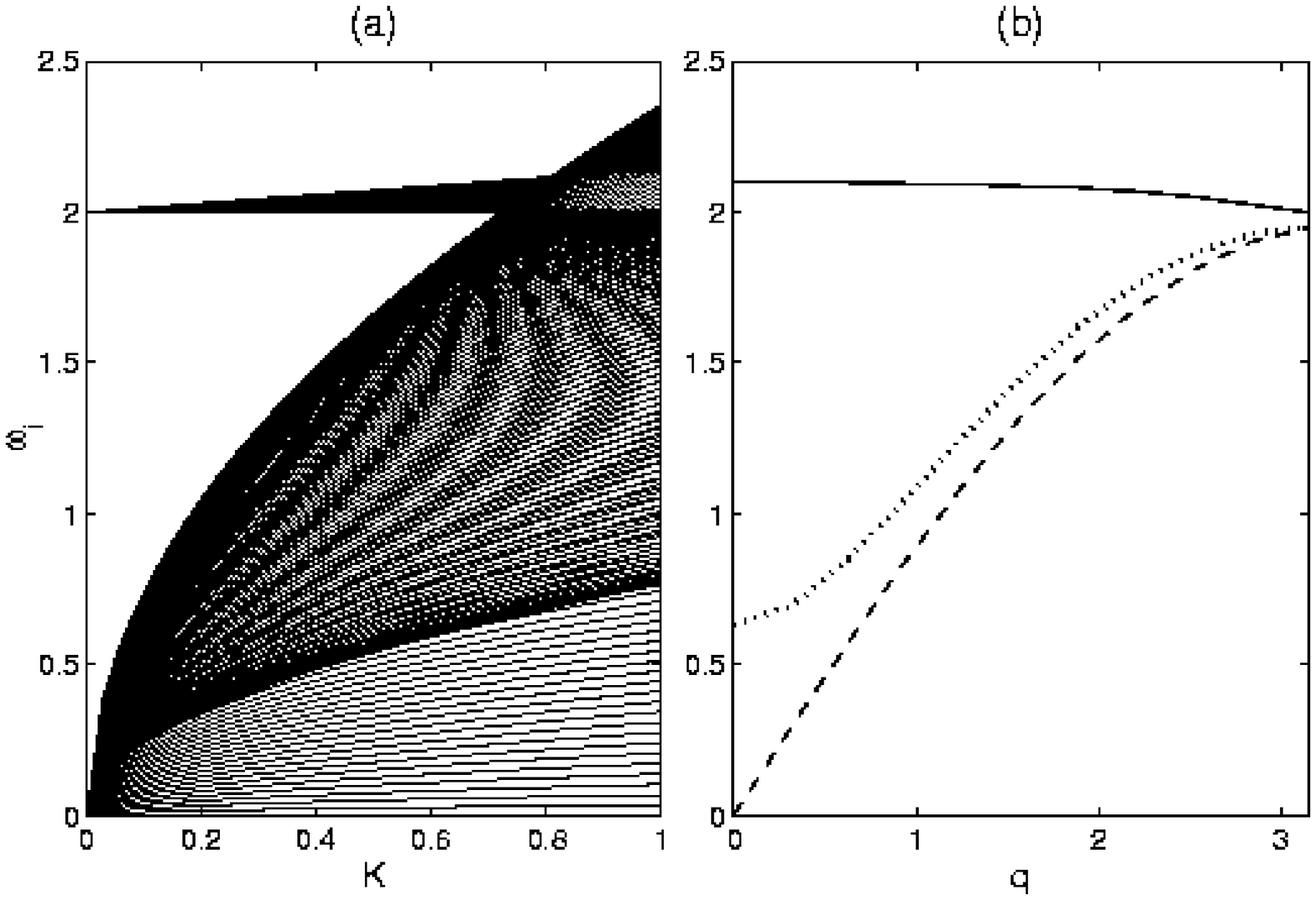}
  \end{center}
  \caption{Spectral analysis of the linear system. (a) Dependence
of the phonon spectrum on the longitudinal elastic constant $K$.
For values of $K$ below $K_c \simeq 0.7$ there exist a gap below
the optical branch where the frequencies of the multi-site
breathers lie. (b) Frequencies of the three phonon branches with
respect to their wave numbers $q$. The $\pi$ modes are the ones on
the vicinity of the multi-site breather frequency. }
  \label{fig:fig10}
\end{figure}

Fig.~\ref{fig:fig10}--a shows  the dependance of the phonon
spectrum on the coupling parameter $K$ and Fig.~\ref{fig:fig10}--b
depicts the three branches for the value of $K=0.683$ used in this
article. It can be seen that there is a phonon gap for values of
$K<K_c$ with $K_c\simeq 0.7$. The Fourier spectrum of the
multi-breathers found consists of a single frequency $\omega_b$
slightly below $\omega_0=2$, the frequency of the isolatedly
vibrating  base pairs, and its harmonics. Furthermore, the modes
in the neighborhood of $\omega_b$ have wave vector $q=\pi$ and
therefore they are not much excited by the perturbation of the
radial variables.

A remarkable fact of the phonon spectrum for $K<K_c$ is the
inversion of the optical spectrum with respect to the planar
PB model \cite{PB}. In the latter, the only variables are
the distances between bases within each base pair, and the
coupling proposed is a standard attractive one, which has
been made more complex afterwards \cite{DPB93,Zhang97}, but without consequences on
the phonon spectrum. The dynamical equations are given by:
\begin{equation}
\ddot{u}_n+V'(u_n)+\epsilon \,(2\,u_n-u_{n-1}-u_{n+1})=0\,,
\label{eq:standard}
\end{equation}
being $V$ the Morse potential
with $V''(0)=\omega_0^2 >0$  and $\epsilon>0$.
The dispersion relation is given by
$\omega^2=\omega_0^2+4\epsilon \sin^2 (q/2)$.
There are some consequences for this system
\cite{MA96,MAF98}: a)  the bottom (linear) mode has $q=0$,
and the top one has $q=\pi$; b) one--site breathers are stable;
c) the tails of a breather or multibreather consists of
in--phase oscillators; d) multibreathers with all the oscillators
in phase (derived from $q=0$) are unstable and with all the
oscillators out of phase (derived from $q=\pi$) are stable
\cite{M97,MJKA02,AACR02,ACSA02}.  If the value of $\epsilon$ were
negative, the conclusions would be reversed.

In the model proposed in this article
considering the helical shape of the DNA molecule,
the optical spectrum is inverted as can be seen in Eqs.~\ref{eq:lincmbp}.
Due to the degeneracies of the model we have not been able to
obtain {\em exact} breathers, but the Fourier spectrum just shows
a frequency below the optical branch and its harmonics, which means that
they are standard breathers in the usual meaning, i.e.,
localized, nonlinear, periodic oscillations. The tails of the multibreathers are,
indeed, of the $q=\pi$ type, according to this inversion. However,
the transcendental consequence arising from the inversion of the effective
type of coupling is that the multibreathers are stable. This is a result
that should not be overlooked: the helical shape of the DNA molecule
provides a means for stable, large amplitude oscillations of base
pairs groups, i.e, for the formation of the denaturation bubble.

\section{Summary}\label{section:summary}

In the present work  we have studied the formation process of broad
H-bridge breathers in DNA molecules.  The double helix formation
 of DNA has been described
in terms of an oscillator model relying on the base-pair picture.
In this context each base pair possesses four degrees of freedom,
namely a radial, an angular one, and two motions for the base pairs
centers of mass. For an extension of the BCP
model \cite{BCP} we have treated the two strands individually
allowing for asymmetric vibrations of the H-bridges. As for the
simplifications of our DNA model, we remark that
 we have not
distinguished between the four different base types and
 therefore, have treated each base as a single entity of fixed mass.
Apart from this, we neglected also the different types of hydrogen
bonds between the two different pairings in DNA, namely the
bridging of the G-C and the A-T pairs by  three and two hydrogen
bonds, respectively.

We concentrate in this paper on the effect of the helical shape of
the DNA molecule, avoiding the introduction of other potential
terms to make it apparent. The result is that this shape brings
about a inversion of the effective type of coupling, with a
inversion of the phonon spectrum, that allows for the existence of
stable multi--site breathers.

We have demonstrated that in the course of the relaxation dynamics
of DNA molecules, which have been displaced in a certain region
from their equilibrium configuration, an equilibrium state is
attained. More precisely, starting from a radially distorted
configuration involving several consecutive base pairs the
twist-opening dynamics relaxes towards a localized state. The
latter is built up from a multi-site H-bridge breather in
combination with a static kink-like profile of the angular
variables connected with the untwist of the helix. Excess energy,
not to be contained in the radial breather, is emitted from the
initially excited region in the form of radial phonons approaching
either ends of the DNA lattice. The relaxation process and the
attainment of a stable breather regime has also been observed for
asymmetric initial conditions when e.g.  bases on only one of the
two strands are displaced  from their equilibrium positions.

We have exploited then the dynamical approach of an equilibrium state for
the construction of breather states of the DNA. To this aim we
have established an iteration scheme, such that, with increasing
steps of applied iterations the dynamics gets closer to an
equilibrium regime supporting a radial breather together with a
kink-like pattern in the angular components. It should be stressed
that the obtained broad H-bridge breathers reproduce realistically
the extension of
 the oscillating bubbles observed in DNA.  However, the almost rectangular shape
of our H-bridge breather solution has to be distinguished from
the  single-site radial breather being reminiscent of an envelope soliton
solution  with a half-width of twenty base pairs obtained in the context of the
BCP model in Ref.\,\,\cite{biophys}.  The multi-site breather with all of its constituents
performing in-phase oscillations of equal amplitude
 renders  all of the involved base pairs accessible to the
functional process (e.g. the transcription) on an equal footing in contrast to the uneven and
less efficient
separation of the two strands associated with the
bell- shaped amplitude pattern when the breather is centered at a single site so that merely
the half width of the exponentially localized profile comprises $15-20$ base pairs.

Our iteration scheme for the construction of H-bridge breathers is
applicable so long as the amplitudes of the radial distortions do
not exceed a critical value.  However, for overcritically large
amplitude (or equivalently, too large amount of  excitation
energy) the dynamics does
 no longer relax onto
a broad H-bridge breather. Remarkably, one rather observes the
directed flow of excitation energy in a single H-bond with the
consequence that this bond may even break up.

Finally, we note that with view to the role of structural
disorder, we have found that the broad  H-bond breathers sustain
 moderate amount of randomness
in the arrangement of the equilibrium positions of the
 bases as well as in the initial
local displacement patterns.

\vspace{0.5cm}
\centerline{\large{\bf Acknowledgments}}

\noindent One of the authors (D.H.)  acknowledges  support by the Deutsche
Forschungsgemeinschaft via a Heisenberg fellowship (He 3049/1-1). J.F.R.A acknowledges
partial support under the LOCNET EU network
HPRN-CT-1999-00163 and D.H. and the
 Institut f\"{u}r Theoretische Physik for their warm hospitality


\newpage

\begin{thebibliography}{99}
\bibitem{Stryer} L. Stryer {\it Biochemistry}, Freeman, New York (1995).
\bibitem{PB} M. Peyrard and A.R. Bishop, Phys. Rev. Lett. {\bf 62},
2755 (1989).
\bibitem{Yakhu} L.V. Yakushevich, Quart. Rev. Biophys.
{\bf 26}, 201 (1993).
\bibitem{Gaeta} G. Gaeta, C. Reiss, M. Peyrard and T. Dauxois, Riv.
Nuovo Cim. {\bf 17}, 1 (1994).
\bibitem{Barbithesis} M. Barbi, {\it Localized Solutions in a
Model of DNA Helicoidal Structure}, PhD Thesis, Universit$\rm
\grave{a}$ degli Studi di Firenze (1998).
\bibitem{BCP} M. Barbi, S. Cocco and M. Peyrard, Phys. Lett. A
{\bf 253}, 358 (1999).
\bibitem{biophys} M. Barbi, S. Cocco, M. Peyrard and S. Ruffo,
J. Biol. Phys. {\bf 24}, 97 (1999).
\bibitem{Cocco99} S. Cocco and R. Monasson, Phys.Rev. Lett.
{\bf 83}(24), 5178 (1999).
\bibitem{Cocco} S. Cocco and R. Monasson, J. Chem. Phys.
{\bf 112}, 10017 (2000).
\bibitem{Agarwal} J. Agarwal and D. Hennig, Physica A.
To appear. (2000).
\bibitem{Campa01} A. Campa, Phys. Rev. E {\bf 63}, 021901 (2001).
To appear. (2000).
\bibitem{Lipniacki} T. Lipniacky, Phys. Rev. E {\bf 64}, 051919
(2001).
\bibitem{Marin} J.L. Mar\'{\i}n
and S. Aubry,
Nonlinearity {\bf 7}, 1623 (1996).
\bibitem{Dauxois} T. Dauxois, M. Peyrard and A.R. Bishop, Phys.
Rev. E {\bf 47}, 684 (1993).
\bibitem{Nyeo} S.L. Nyeo and I.C. Yang, Mod. Phys. Lett. {\bf 14},
313 (2000).
\bibitem{Theodorakopoulos} N. Theodorakopoulos, T. Dauxois and M.
Peyrard, Phys. Rev. Lett. {\bf 85}, 6 (2000).
\bibitem{Marko} J.F. Marko and E.D. Siggia, Science {\bf 265}, 506
(1994); T.R. Strick, J.F. Allemand, D. Bensimon, A.
Bensimon, and V. Croquette, Science {\bf 271}, 1835 (1996).
\bibitem{Smith} S.B. Smith, L. Finzi and C. Bustamante, Science
{\bf 258}, 1122 (1992).
\bibitem{Essevaz} B. Essevaz-Roulet, U. Bockelmann and F. Heslot,
Proc. Natl. Acad. Sci. U.S.A. {\bf 94}, 11935 (1197).
\bibitem{Bockelmann} U. Bockelmann, B. Essevaz-Roulet, and F.
Heslot, Phys. Rev. Lett. {\bf 79}, 4489 (1997); Phys. Rev. E {\bf
58}, 2386 (1998).
\bibitem{Clausen} H. Clausen-Schaumann, M. Rief, C. Tolksdorf,
H.E. Gaub, Biophys. J. {\bf 78}, 1997 (2000).
\bibitem{Crisona} N.J. Crisona, T.R. Strick, D. Bensimon, V.
Croquette and N.R. Cozzarelli, Genes $\&$ Dev. {\bf 14}, 2881
(2000).
\bibitem{Ye} Y.-J. Ye, R.-S. Chen, A. Martinez, P. Otto and
J. Ladik, Sol. Stat. Comm. {\bf 112}, 139 (1999).
\bibitem{Moroz} J.D. Moroz and P. Nelson, Proc. Natl. Acad. Sci.
U.S.A. {\bf 94}, 14418 (1997).
\bibitem{Prohofsky} E.W. Prohofsky, K.C. Lu, L.L van Zandt and B.F. Putnam,
Phys. Lett. A {\bf 70}, 492 (1979).
\bibitem{Flach} S. Flach and C.R. Willis, Physics Reports {\bf
295}, 181 (1998).
\bibitem{DPB93} T. Dauxois, M. Peyrard and A.R. Bishop, Phys. Rev. E {\bf 47}(1), R44 (1993).
\bibitem{Zhang97} Y. Zhang, W. Zheng, J. Liu and Y. Z. Chen, Phys. Rev. E {\bf 56}(6),
7100 (1997).
\bibitem{MA96}J.L.  Mar{\'{\i}}n and S. Aubry, Nonlinearity {\bf 9}, 1501 (1996).
\bibitem{MAF98}J.L.  Mar{\'{\i}}n, S. Aubry and LM Flor\'{\i}a, Physica D {\bf 113}, 283 (1998).
\bibitem{M97}J.L.  Mar{\'{\i}}n, Ph. {D}. dissertation, Universidad de {Z}aragoza, 1997.
\bibitem{A97}S. Aubry, Physica D {\bf 103}, 201 (1997).
\bibitem{MJKA02}A.M. Morgante, M. Johansson, G. Kopidakis and S. Aubry, Physica D {\bf 162}, 53 (2002).
\bibitem{AACR02}A. Alvarez, J.F.R. Archilla, J. Cuevas and F.R. Romero,
New Journal of Physics {\bf 4}, 72.1 (2002).
\bibitem{ACSA02}J.F.R. Archilla, J. Cuevas, B. S\'anchez--Rey and A. Alvarez,
 Physica D {\bf 180}, 235 (2003).

\end{thebibliography}
\end{document}